\documentclass[11pt,english,twocolumn,10pt,pre]{revtex4-1}
\usepackage[T1]{fontenc}
\usepackage[latin9]{inputenc}
\usepackage{geometry}
\usepackage{verbatim}
\usepackage{amsmath}
\usepackage{graphicx}

\makeatletter
\usepackage[usenames]{color}

\usepackage{hyperref}

\makeatother

\usepackage{babel}
\begin{document}

\title{Hypothesis testing of scientific Monte Carlo calculations}

\author{Markus Wallerberger, Emanuel Gull}

\affiliation{Department of Physics, University of Michigan, Ann Arbor, MI 48109}
\begin{abstract}
The steadily increasing size of scientific Monte Carlo simulations
and the desire for robust, correct, and reproducible results necessitates
rigorous testing procedures for scientific simulations in order to
detect numerical problems and programming bugs. However, the testing
paradigms developed for deterministic algorithms have proven to be
ill suited for stochastic algorithms. In this paper we demonstrate
explicitly how the technique of statistical hypothesis testing, which
is in wide use in other fields of science, can be used to devise automatic
and reliable tests for Monte Carlo methods, and we show that these
tests are able to detect some of the common problems encountered in
stochastic scientific simulations. We argue that hypothesis testing
should become part of the standard testing toolkit for scientific
simulations.
\end{abstract}
\maketitle

\section{Introduction}

Scientific computing, i.e., the process of obtaining numerical results
from scientific theories using algorithms, relies on correct and reproducible
implementations of computer programs. In condensed matter and statistical
physics, these computer programs were traditionally small, often implemented
by a single researcher, and tested and debugged by hand until no more
problems could be found.

Over time, the size and complexity of programs in this field has grown
rapidly. For example, computer programs for complex many-body problems,
such as finding the ground state energy of an interacting solid \cite{QMCPACK12}
or evaluating response functions of correlated quantum impurity models
\cite{ALPSCore17,TRIQS15}, now span hundreds of thousands of lines
that are developed and maintained by large and constantly changing
teams. For such programs, manual testing becomes inefficient and expensive.

This challenge is not unique to scientific computing, and software
engineering has responded by establishing automated testing practices.
The corresponding arsenal of methods includes, in order of increasing
granularity: contract programming, where invariants in the program
state are verified continuously during execution \cite{meyer-comp-1992};
unit tests, which ensure the correctness of small sections of the
code \cite{binder-testing-2000}; as well as integration and system
tests, which check that implementations yield correct non-trivial
results for predefined benchmark problems \cite{ould-testing-1986}. 

These techniques have permeated scientific software engineering \cite{dubois-cse05},
and they are by now standard in many computational science packages.
Combined with continuous testing, i.e., the automatic execution of
tests after a change to the code base, they have led to a massive
improvement of the quality and resilience of scientific software \cite{sanders-ieeesw08}.

Nevertheless, there is a large part of computational and statistical
physics where such tests were so far not practical, namely the field
of stochastic Monte Carlo simulations. In this domain, results make
use of random or pseudo-random number generators, and are therefore
intrinsically stochastic in nature.  Agreement with a reference result
has to be ``within error bars'' only.

As far as we are aware, most practitioners of these techniques therefore
either enforce a deterministic procedure (e.g., a simulation with
a fixed seed of the pseudo-random number generator or an otherwise
fixed sequence of updates on a given configuration) or resort to ``visual
inspection'' of the results to determine agreement between simulation
and reference, neither of which is optimal. The former breaks whenever
the sequence or ratio of updates are changed, and therefore is prone
to false negatives, i.e., failed tests even though the results are
correct. The latter relies on human intervention and is therefore
neither reliable nor automatable.

In this paper, we show how tools of statistics \cite{arnold-stat2a},
known for more than a century and in wide use in many fields, should
be used to construct automated tests for physics simulations. Our
formulations are general and applicable to any stochastic simulation.
While we are not aware of applications to physics so far, we emphasize
that similar applications have been pioneered both in the field of
image synthesis \cite{sevcikova-psta-2006} and urban simulations
\cite{subr-pcga-2007}.

In the remainder of this paper we will introduce the concept of statistical
testing or ``hypothesis testing'' in Sec. \ref{sec:scalar} and
\ref{sec:series}, with applications to the two-dimensional Ising
model. Sec. \ref{sec:AIM} shows an application to the Anderson impurity
model, and Sec.~\ref{sec:conclusions} summarizes our conclusions.

\section{Scalar Tests\label{sec:scalar}}

\subsection{One-sample test for the mean\label{sec:1spl}}

The basic idea of statistical hypothesis testing in the context of
Monte Carlo is straight-forward: one first chooses a model for which
an exact benchmark result $y$ exists. The null hypothesis, $H_{0}$,
is that there is no significant difference between this reference
result and the expectation value $\mathrm{E}[\hat{X}]$ of a simulation
with the estimator $\hat{X}$ \cite{subr-pcga-2007}. The alternate
hypothesis, $H_{1}$, is that this is not the case:\begin{subequations}
\begin{align}
H_{0} & :\mathrm{E}[\hat{X}]=y\label{eq:h0}\\
H_{1} & :\mathrm{E}[\hat{X}]\neq y.\label{eq:h1}
\end{align}
\label{eqs:h01}\end{subequations}

We first discuss the scalar case. Let $\hat{X}$ be a ``simple''
Monte Carlo estimator, i.e., an average $\langle X\rangle$ over $N$
independent random variables identically distributed according to
$X$. (In the case of sampling on a Markov chain, one has to correct
the number $N^{\prime}$ of Monte Carlo samples by the integrated
autocorrelation time: $N=N^{\prime}/\tau_{\mathrm{int},X}$.) We then
find:
\begin{equation}
\frac{\langle X\rangle-y}{\sigma_{X}/\sqrt{N}}\sim t_{N-1},\label{eq:t}
\end{equation}
where $\sim$ is shorthand for ``is distributed according to'',
$t_{\nu}$ is Student's $t$ distribution for $\nu$ degrees of freedom
and $\sigma_{X}^{2}$ is the variance of $X$.

Following standard practice \cite{neyman-ptrsoca-1933}, we turn Eq.~(\ref{eq:t})
into a likelihood estimate for $H_{0}$, known as Student's $t$ test.
We compute the two-sided $p$-value as $p=2P^{-1}(-|z|)$, where $P^{-1}$
is the inverse of the cumulative distribution function of the right-hand
side and $z$ is the observed left-hand side in Eq.~(\ref{eq:t}).
Finally, we compare $p$ with a significance level $\alpha\in(0,1)$
and reject the null hypothesis (\ref{eq:h0}) if $p<\alpha$. In other
words, the $p$ value is the probability of observing $z$ or a ``more
unlikely'' event given $H_{0}$, and we reject the $H_{0}$ if that
probability becomes smaller than $\alpha$.

\begin{figure}
\includegraphics[width=1\columnwidth]{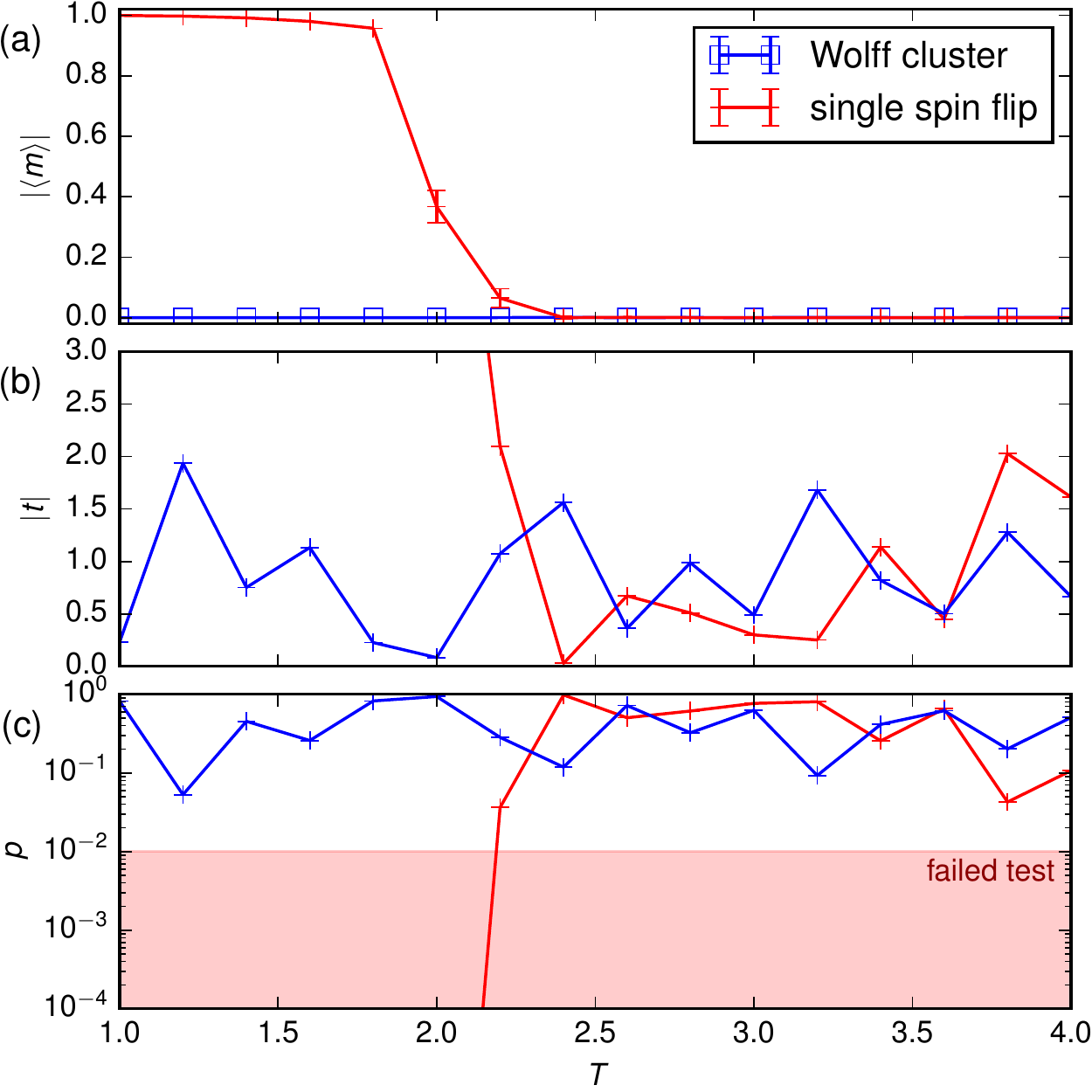}

\caption{Scalar one-sample test against $\langle m\rangle=0$ for single spin-flip
updates (red curves) and Wolff cluster updates (blue curves) in a
classical two-dimensional Ising model with length $L=16$: (a) result
for $\langle m\rangle$ from $N^{\prime}=10^{6}$ Monte Carlo sweeps
(a sweep is either a set of $L^{2}$ single spin flips or a single
cluster update); (b) $|t|$ score as the left-hand side of Eq.~\ref{eq:t};
and (c) $p$ values from a two-tailed test of the $t$ score against
the Student $t_{N-1}$ distribution (the shaded area indicates $p<\alpha=0.01$
and thus a failed test).}

\label{fig:m}
\end{figure}

Let us illustrate the procedure with a simple example, the ferromagnetic
Ising model \cite{LandauBinder05}
\begin{equation}
\mathcal{H}=-\sum_{\langle ij\rangle}\sigma_{i}\sigma_{j},\label{eq:ising}
\end{equation}
where $\langle ij\rangle$ runs over all pairs of directly neighboring
Ising spins $\sigma_{i}\in\{1,-1\}$ on a $L\times L$ square lattice
with periodic boundary conditions and $L=16$. Since the system is
finite and there is no external magnetic field, $\langle m\rangle=0$.
We perform a Markov chain Monte Carlo simulation \cite{Metropolis53}
for Eq.~(\ref{eq:ising}) for two different types of updates: (a)
a set of single spin flips $\sigma_{i}\to-\sigma_{i}$, and (b) Wolff
cluster updates \cite{Wolff89}. In both cases, the magnetization
estimator is constructed as $\hat{m}=\langle\sum_{i}\sigma_{i}\rangle/L^{2}$.

Fig.~\ref{fig:m} shows the temperature-dependent magnetization curves
obtained by the simulation. From Fig.~\ref{fig:m}(a), we immediately
see that the single spin flip updates (red curve) produce a spurious
spin polarization at low temperature for the parameters chosen. This
is to be expected, since in order to restore $\langle m\rangle=0$,
all spins must be flipped, which due to the exponentially divergent
autocorrelation time $\tau\propto\exp(L)$ requires far more updates
than performed in our test. Figure~\ref{fig:m}(b) shows the $|t|$
score or deviation in units of the standard error computed from Eq.~\ref{eq:t}.
Figure~\ref{fig:m}(c) shows the $p$ value as result of a two-tailed
test with the Student distribution, which amounts to $p=2P^{-1}(-|t|)$.
If we choose a significance level of $\alpha=0.01$, we see that the
test fails for all temperatures below the critical temperature, $T<2.2$.
In contrast, the Wolff updates, which circumvent the problem of divergent
autocorrelation times by updating clusters of spins, pass the test
for all temperatures.

The spurious spin polarization is already obvious from a fleeting
inspection of Fig.~\ref{fig:m}(a), and a formal verification of
Eq.~(\ref{eq:h1}) may seem superfluous. However, we emphasize that
the formal procedure can easily be turned into an automated test and
run as part of an automated test suite. This extends the test coverage
from the deterministic parts of the algorithm to the stochastic updates
and the magnetization estimator and its autocorrelation effects.

The choice of significance level $\alpha$ is a trade-off between
the probability of two kinds of errors:\begin{subequations}
\begin{align}
\alpha & =P(H_{0}\mathrm{\,rejected}\,|\,H_{0}\mathrm{\,is\,true})\label{eq:alpha}\\
\beta & =P(H_{0}\mathrm{\,accepted}\,|\,H_{0}\mathrm{\,is\,false}),\label{eq:beta}
\end{align}
\end{subequations}known as type-I and type-II errors, or false positives
and false negatives, respectively. We empirically find that the rather
conservative $\alpha\approx0.01$ provides such a good trade-off for
a single test. In the case of a test suite of $K$ tests, one can
either substitute $\alpha\to\alpha^{\prime}\approx\alpha/K$ to keep
the probability of a type-I error constant or keep the threshold as-is
to keep the probability of a type-II error constant. The former scheme
is suited for automatized stochastic unit tests, the latter strategy
is advantageous when combined with test refinement. In such a scheme,
we choose a window $p\in[\alpha,\beta)$ corresponding to ambiguous
test results and re-run these cases with double the number of samples
until they are either accepted or rejected.

\subsection{Two-sample test; biased estimator\label{sec:2spl}}

In many cases, exact benchmark results may not be available or cumbersome
to obtain. In these cases, we can also compare two stochastic results:
the estimator $\hat{X}$ to be tested and a trusted estimator $\hat{Y}$.
This corresponds to replacing $y$ with $\mathrm{E}[\hat{Y}]$ in
Eqs.~\ref{eqs:h01}:\begin{subequations}
\begin{align}
H_{0} & :\mathrm{E}[\hat{X}]=\mathrm{E}[\hat{Y}]\label{eq:h20}\\
H_{1} & :\mathrm{E}[\hat{X}]\neq\mathrm{E}[\hat{Y}]\label{eq:h21}
\end{align}
\label{eqs:h201}\end{subequations}In the scalar case, with $\hat{X}$
and $\hat{Y}$ averages over $N_{X}$ and $N_{Y}$ independent random
variables distributed according to $X$ and $Y$, we find the analogue
of Eq.~(\ref{eq:t}), 
\begin{equation}
\frac{\langle X\rangle-\langle Y\rangle}{\sigma/N_{\mu}}\sim t_{N_{X}+N_{Y}-2},\label{eq:tpooled}
\end{equation}
with $N_{\mu}^{-1}=N_{X}^{-1}+N_{Y}^{-1}$ and the pooled variance
\begin{equation}
\sigma^{2}=\frac{(N_{X}-1)\sigma_{X}^{2}+(N_{Y}-1)\sigma_{Y}^{2}}{N_{X}+N_{Y}-2}.\label{eq:pooled}
\end{equation}
The rest of the test proceeds exactly the same as for the one-sample
test (Sec.~\ref{sec:1spl}).

\begin{figure}
\includegraphics[width=1\columnwidth]{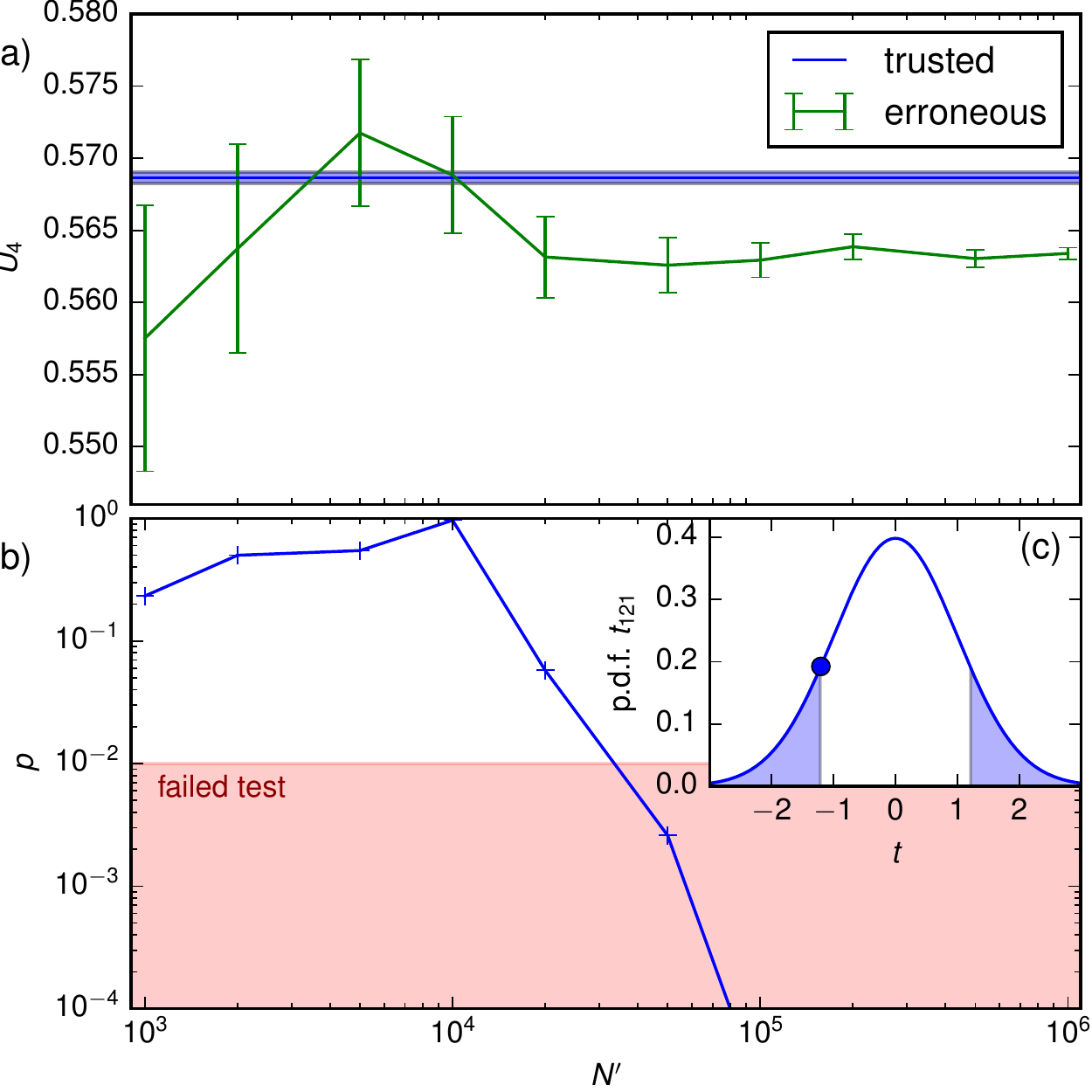}

\caption{Scalar two-sample test for the Binder cumulant $U_{4}$ in the Ising
model obtained from Markov chain Monte Carlo with Wolff cluster updates
at temperature $T=2.3$ for system length $L=32$. (a) Sample mean
of $U_{4}$ of an erroneous implementation for different simulation
times $N$, and the result of a correct reference simulation; (b)
corresponding $p$ values computed using Eq.~(\ref{eq:tpooled});
(c) illustration of the $p$ value for $N^{\prime}=10^{3}$ ($N=123$
after the removal of autocorrelation and variance pooling) as the
shaded area under the probability density function (p.d.f.) of the
corresponding $t$ distribution.}

\label{fig:u4}
\end{figure}

As an example, we reexamine data from the Ising model, this time on
a $32\times32$ square lattice. We verify the estimator for the Binder
cumulant \cite{Binder81}
\begin{equation}
\hat{U}_{4}=\frac{\langle m^{4}\rangle}{1-3\langle m^{2}\rangle^{2}}.\label{eq:u4}
\end{equation}
The Student $t^{2}$ test is sensitive to non-Gaussian distributed
errors, which occur in the computation of Eq.~(\ref{eq:u4}) due
to non-linear error propagation. To remedy this, we use the jackknife
resampling procedure, which replaces Eq.~(\ref{eq:u4}) with a simple
average $\langle U_{4}^{\prime}\rangle$ over pseudovalues $U_{4}^{\prime}$,
removing the linear order of the bias and restoring the validity of
Eq.~(\ref{eq:tpooled}) \cite{efron-jackknife}. Alternatively, one
could abandon the the Student test altogether in favor of the parametric
bootstrap method \cite{efron-jackknife}. However, we will see that
the jackknife method suffices in our case.

To simulate a common programming error, we have artificially broken
periodic boundary conditions on the corners of the lattice (they are
reduced to having two neighbors each). Fig.~\ref{fig:u4}(a) compares
this erroneous implementation (green curve) with a simulation result
where the error is not present. As evident from Fig.~\ref{fig:u4}(b),
as we increase the number of Monte Carlo sweeps $N^{\prime}$, the
error bars shrink and the null hypothesis (\ref{eq:h20}) is rejected
more and more strongly.

\section{Data series tests\label{sec:series}}

\subsection{Tests for the mean}

\begin{figure}
\includegraphics[width=1\columnwidth]{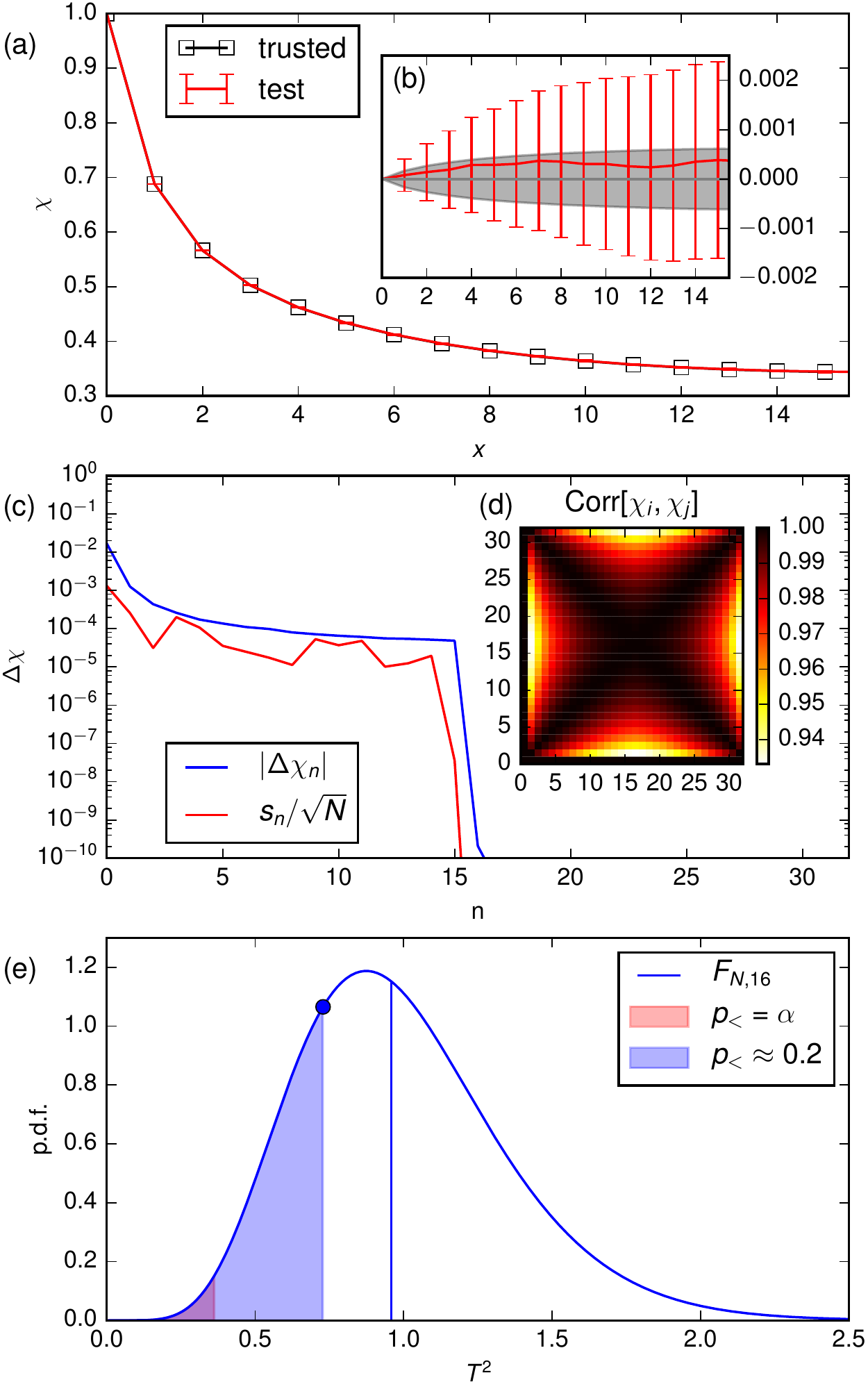}

\caption{Vector-valued two-sample test on the spin correlation function $\chi_{x,0}$
(cf.~Eq.~(\ref{eq:chi})). (a) Simulation result for Wolff updates
(black curve) and single spin-flip updates (red curve); (b) deviation
of spin-flip from Wolff update (the shaded region are the Wolff result
error bars); (c) Projected deviations and errors (numerator and denominator
in the l.h.s. of Eq.~(\ref{eq:frestr})); (d) correlation matrix
(\ref{eq:corr}); (e) p.d.f. of the corresponding $F$ distribution
in Eq.~(\ref{eq:frestr}) with the mean as vertical blue line and
$p_{<}$-value as blue shaded area to the left of the observed $T^{2}$
score (blue dot) as well as test failure threshold as red-shaded area.}

\label{fig:chi}
\end{figure}

While tests for scalar quantities (Sec.~\ref{sec:scalar}) are useful,
we empirically find that it is often easier to identify problems when
comparing functions and data series. In the case of a one-sample test,
this corresponds to the benchmark result $y$ being a vector of $n$
elements rather than a scalar. Consequently, the Monte Carlo estimator
$\hat{X}$ is vector-valued. Again assuming independent and identically
distributed results, Eq.~(\ref{eq:t}) is replaced by \cite{hotelling-amstat-1931}
\begin{equation}
N(\langle X\rangle-x_{0})^{\mathrm{T}}\Sigma_{X}^{-1}(\langle X\rangle-x_{0})\sim\frac{n(N-1)}{N-n}F_{n,N-n}\;,\label{eq:f}
\end{equation}
where $\langle X\rangle$ is the sample mean, $\Sigma_{X}$ is the
sample covariance matrix, and $F_{a,b}$ is the Fisher\textendash Snedecor
distribution with parameters $a$, $b$. One proceeds in a similar
way to the Student's $t$-test. The observed left-hand side of Eq.~(\ref{eq:f})
is again used as the test statistic and checked against the right-hand
side distribution. However, since the $F$ distribution is not symmetric
for low $n$ (cf.~Fig.~\ref{fig:chi}(c)), one uses two one-sided
tests instead of a two-sided test and subsequently obtains two $p$-values,
which we will call $p_{<}$ and $p_{>}$. This is known as Hotelling's
$T^{2}$ test.

In the case where we compare the estimator to a trusted result $\langle Y\rangle$,
we proceed similar as in Sec.~\ref{sec:2spl} and replace Eq.~(\ref{eq:f})
with:
\begin{equation}
\begin{split} & N_{\mu}(\langle X\rangle-\langle Y\rangle)^{\mathrm{T}}\Sigma^{-1}(\langle X\rangle-\langle Y\rangle)\\
 & \qquad\sim\frac{n(N_{X}+N_{Y}-2)}{N_{X}+N_{Y}-n-1}F_{n,N_{X}+N_{Y}-n-1}\;,
\end{split}
\label{eq:fpooled}
\end{equation}
where $\Sigma$ is the pooled covariance obtained by replacing all
variances $\sigma_{a}^{2}$ with covariance matrices $\Sigma_{a}$
in Eq.~(\ref{eq:pooled}).

In order to illustrate the procedure, we revisit our Ising model example
for $L=32$ and $T=2.3$ (close to the critical temperature) and examine
the spin correlation function
\begin{equation}
\begin{split}\chi_{x,y} & =\langle\sigma_{0,0}\sigma_{x,y}\rangle\\
 & =\frac{1}{L^{2}}\langle\sum_{k,q}\sum_{x^{\prime},y^{\prime}}\mathcal{F}_{x,y;k,q}^{-1}|\mathcal{F}_{k,q;x^{\prime},y^{\prime}}\sigma_{x^{\prime},y^{\prime}}|^{2}\rangle,
\end{split}
\label{eq:chi}
\end{equation}
where $(x,y)$ denote row and column of the lattice site, and $\mathcal{F}$
denotes the discrete Fourier transform used in the actual estimator.
Fig.~\ref{fig:chi}(a) shows $\chi_{x0}$ for the Wolff cluster update
(black curve), which we take as the trusted result, and for a set
of spin-flip updates (red curve). The inset Fig.~\ref{fig:chi}(b)
shows the deviation of the red curve from the black one, where the
shaded region marks the error bars of the Wolff update result. We
see significant correlation of the error bars, which underscores the
importance of a proper treatment of the covariance matrix (cf.~Fig.~\ref{fig:chi}(d)).

\subsection{Cross-correlated data}

A common complication with the $T^{2}$ test are perfect correlation
or anti-correlation within the dataset (duplicates), which implies
a singular covariance matrix in Eq.~(\ref{eq:f}). In our example,
the symmetry of the system implies $\chi_{x,y}=x_{L-x,y}$, thus half
of the points yielded by the estimator (\ref{eq:chi}) are just copies
of the other half. We can confirm this by examining the correlation
matrix:
\begin{equation}
\mathrm{Corr}[\chi_{x,0},\chi_{x^{\prime},0}]=\frac{\mathrm{Cov}[\chi_{x,0},\chi_{x^{\prime},0}]}{\sqrt{\mathrm{Var}[\chi_{x,0}]\mathrm{Var}[\chi_{x^{\prime},0}]}},\label{eq:corr}
\end{equation}
plotted in Fig.~\ref{fig:chi}(d), which is one on the anti-diagonal.

This can be solved by first diagonalizing $\Sigma$ and retaining
only the non-zero eigenvalues (a relative threshold of $10^{-14}$
seems to be practical for most cases we studied):
\begin{equation}
\Sigma=\mathcal{P}\;\mathrm{diag}(s_{1}^{2},\ldots,s_{m}^{2})\;\mathcal{P}^{\mathrm{T}},\label{eq:restr}
\end{equation}
where $\mathcal{P}$ is the $n\times m$ projection to the non-zero
eigenvalues $s_{1}^{2},\ldots,s_{m}^{2}$. This is shown in Fig.~\ref{fig:chi}(c),
where there is sharp drop of $s_{n}$ (red curve) in magnitude after
$m=15$. Eq.~(\ref{eq:f}) is then amended to:
\begin{equation}
\sum_{i=1}^{m}\frac{|\sum_{k=1}^{n}\mathcal{P}_{ki}(\langle X_{k}\rangle-y_{k})|^{2}}{s_{i}^{2}/N}\sim\frac{m(N-1)}{N-m}F_{m,N-m}.\label{eq:frestr}
\end{equation}
Note the reduction in the degrees of freedom from $n$ to $m$, which
corresponds to discarding the $n-m$ correlated data points. Note
also that for $n=m$ Eq.~(\ref{eq:f}) and Eq.~(\ref{eq:frestr})
are equivalent, such that in practical calculations, one can always
use Eq.~(\ref{eq:frestr}). Finally, we perform a $T^{2}$ test against
the appropriate $F$ distribution and find that the null hypothesis
is accepted with $p\approx0.2$ (Fig.~\ref{fig:chi}(e)).

\subsection{Error bars}

By using the sample mean and covariance as input rather than the individual
samples, one can interpret Hotelling's $t^{2}$ test as statistical
test on the error bars $\sigma_{0}$ :\begin{subequations}
\begin{align}
H_{0} & :\sigma=\sigma_{0}\label{eq:h0err}\\
H_{1}^{-} & :\sigma<\sigma_{0}\label{eq:h1m}\\
H_{1}^{+} & :\sigma>\sigma_{0}\label{eq:h1p}
\end{align}
\label{eqs:h011}\end{subequations}For a scalar estimator (Sec.~\ref{sec:scalar}),
we can distinguish $H_{0}$ from $H_{1}^{-}$: error bars being ``too
small'' (\ref{eq:h1m}) is equivalent to the result being inconsistent
with the benchmark (Eq.~(\ref{eq:h1})). However, we cannot test
against $H_{1}^{+}$, since we may have accidentally hit the benchmark
accurately. Using a data series, we can also distinguish it from $H_{1}^{+}$,
formalizing the rule that ``roughly two-thirds of the data should
fall within one-sigma error-bars''. This is reflected in the fact
that for $n>1$, the $F$ distribution turns from a one-tailed to
a two-tailed distribution, and becomes more symmetric around $1$
as $n$ gets larger. We can make use of this by testing the lower
tail as score for $H_{1}^{+}$ and the upper tail as score for $H_{1}^{-}$.

This procedure is illustrated at the example of the estimator for
$\chi_{i,0}$ (Sec.~\ref{sec:series}). If we ignore the cross-correlation
(Fig.~\ref{fig:chi}d) and interpret the error bars in Fig.~\ref{fig:chi}b
as uncorrelated errors, it is evident from visual inspection that
they are too large. We can confirm this numerically by (erroneously)
plugging the diagonal elements $\Sigma_{ii}$ of the covariance matrix
instead of its eigenvalues $s_{i}$ into Eq.~(\ref{eq:frestr}).
We then find a $T^{2}$ score of $0.03$ and an acceptance of the
lower alternate hypothesis $H_{1}^{+}$ (\ref{eq:h1m}) with $p=1-10^{-16}$.

\section{Example: Anderson impurity model\label{sec:AIM}}

\begin{figure}
\includegraphics[width=1\columnwidth]{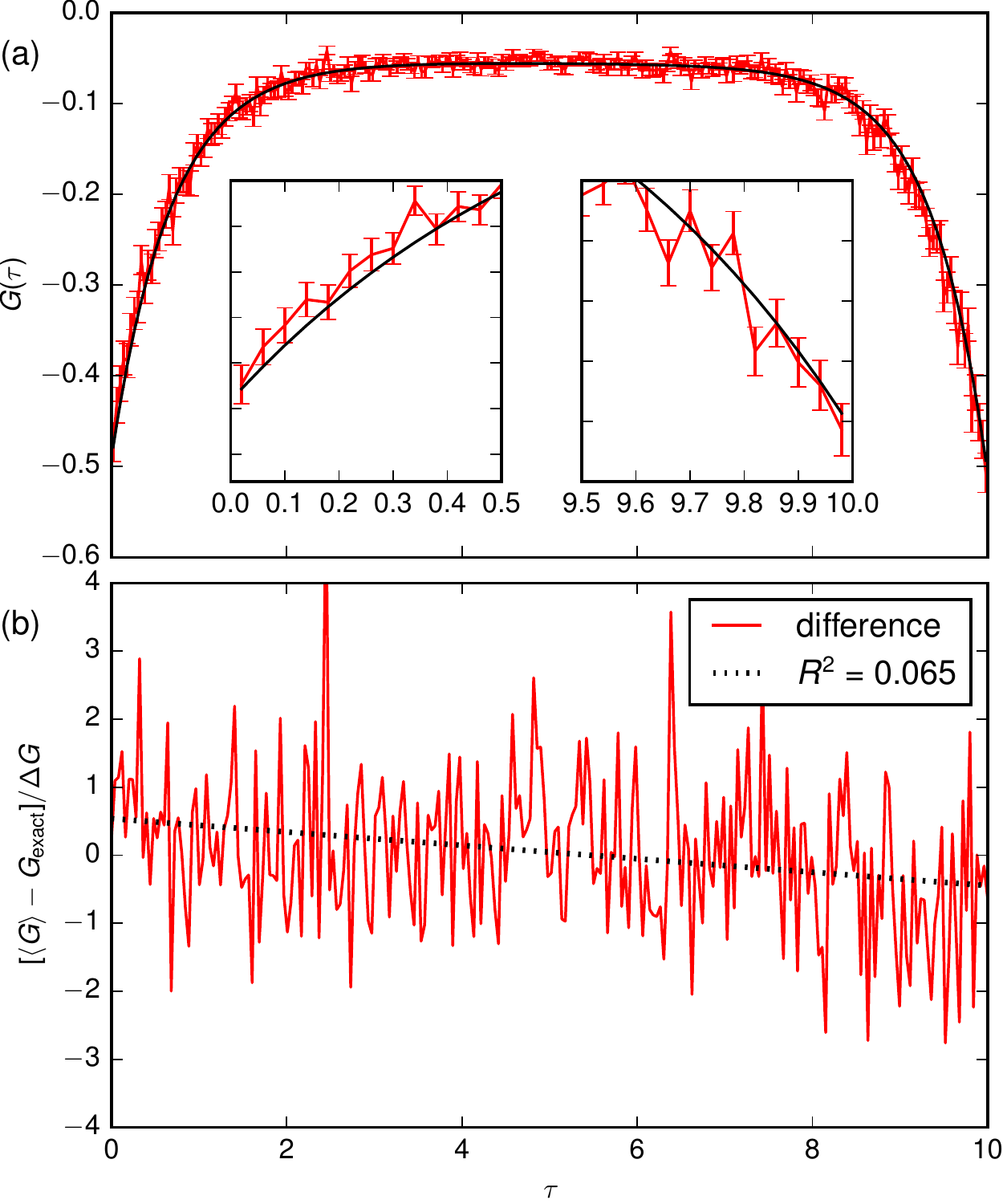}

\caption{(a) Green's function $G(\tau)$ for AIM parameters as in main text:
Monte Carlo result (red) and exact result with an artificially introduced
shift of the half bin size, i.e., $G(\tau-0.02)$, (black) modeling
a binning error; (b) deviation from the exact result in multiples
of the standard error as well as a linear regression (black dashed
line) $0.539-0.098\tau$ with a goodness of fit of $R^{2}\approx0.065$.}
\label{fig:gf}
\end{figure}

To illustrate our method on a research example, we examine the single-orbital
Anderson impurity model \cite{Anderson61} (AIM) which characterizes
a few discrete and potentially correlated impurity states coupled
to a non-interacting bath. The model is in wide use in nano- and transport
science \cite{Hanson07,Brako81} and as an auxiliary model in the
dynamical mean field theory \cite{georges-rmp-1996}, and in many
parameter regimes quantum Monte Carlo methods are the standard tools
for obtaining its properties \cite{gull-rmp-2011}. Its Hamiltonian
is
\begin{equation}
\begin{split}\mathcal{H}= & \;Uc_{\uparrow}^{\dagger}c_{\downarrow}^{\dagger}c_{\downarrow}c_{\uparrow}-\mu\sum_{\sigma}c_{\sigma}^{\dagger}c_{\sigma}\\
 & +\sum_{p\sigma}(V_{p\sigma}f_{p\sigma}^{\dagger}c_{\sigma}+\mathrm{h.c.})+\sum_{p\sigma}\epsilon_{p}f_{p\sigma}^{\dagger}f_{p\sigma}.
\end{split}
\label{eq:aim}
\end{equation}
Here, $c_{\sigma}$ annihilates a fermion of spin-$\sigma$ on the
impurity and $f_{p\sigma}$ annihilates a bath fermion of momentum
$p$ and spin $\sigma$. Impurity interactions are characterized by
$U,$ $\mu$ denotes a chemical potential, $V$ a spin- and momentum
dependent hybridization term, and $\epsilon_{p}$ a momentum-dependent
bath dispersion. In the context of the AIM, a truncation of the bath
to a few states and subsequent exact diagonalization of the finite
system is particularly suitable for testing. While the complexity
of solving the model with Monte Carlo methods is the same as for a
model without bath truncation, one empirically finds that the truncated
model shares much of the physics of the AIM\emph{ }and can thus be
used to generate non-trivial, analytically accessible test cases for
Monte Carlo simulations.

Our example consists of two momenta and correspondingly two bath sites
with energies of $\epsilon_{p}=\pm0.5$ and a hybridization strength
$V=1$, as well as $U=5$, $\mu=U/2$, and temperature $T=1/10$.
Stochastic results were obtained using continuous-time quantum Monte
Carlo in the hybridization expansion \cite{werner-prb-2006,gull-rmp-2011}.

The imaginary time Green's function $G(\tau)=-\langle Tc(\tau)c^{\dagger}(0)\rangle,$
which is the fundamental quantity of interest in this model and which
is directly related to the interacting spectral function, is shown
in Fig.~\ref{fig:gf}. In order to mimic the effect of a typical
binning programming error, we have shifted the exact result (black)
by half a bin $G(\tau)\to G(\tau-0.02)$. The top panel shows that
the Monte Carlo result (red) is still consistent with the exact result
in this case when gauged by visual inspection. This is reinforced
by the bottom panel, where the deviation from the exact result in
multiples of the standard error is plotted (red). Overall we find
the expected result, even though a linear fit of the data (shown as
black dashed line) shows a slight downward slope indicative of a problem.

However, Hotelling's $T^{2}$ test finds a test statistic of $T^{2}\approx1.28$
and therefore a rejection of the null hypothesis in favor of $H_{1}^{+}$
with $p\approx0.0026$ (about three sigma). This is because by using
all $n=250$ data points, the test becomes sensitive to a small increase
of the values outside of error bars. Systematically increasing the
statistics would eventually expose the deviation to visual inspection.

\section{Conclusions\label{sec:conclusions}}

In this paper, we have shown how hypothesis testing can be used to
develop tests for code correctness of Monte Carlo codes in statistical
and condensed matter physics. We also have shown how these tests are
sensitive to different types of simulation problems, and how they
can therefore be used as diagnostic tools to ensure the correctness
of simulations.

The mathematical framework for hypothesis testing has been known for
over 100 years and statistical tests are in wide use across many scientific
fields. Despite this, the technique is not used on a routine basis
for testing scientific simulation results. With the advent of automatic
testing and unit test frameworks, which have permeated most of software
engineering and to some extent also scientific computing, our techniques
add to the testing toolkits that can be used to systematically ensure
correctness and reproducibility of stochastic physics simulations.
These tests integrate well into existing testing frameworks and can
validate parts of the programs that are otherwise difficult to test.

Hypothesis testing allows to gain and keep trust in complex codes
as they undergo modifications, and to uncover problems that are difficult
to uncover by other means, e.g. manual visual inspection. This both
increases the speed of scientific software development and the trust
in results produced by complex computer programs.

In our opinion hypothesis testing should be widely adopted in statistical
simulation codes and should become a standard tool in scientific software
development. While implementing these tests carries a small overhead,
we argue that rigorous, frequent, and automatic testing is necessary
for today's codes, especially in light of the replication crisis \cite{Baker16}
observed in other fields of science.

The code for the stochastic solvers and the hypothesis testing post-processing
scripts are available from the authors upon request. An open-source
software implementation of hypothesis testing is scheduled for inclusion
in the upcoming version of the ALPS core libraries.\cite{ALPSCore17}
\begin{acknowledgments}
The authors would like to thank Alexander Gaenko for fruitful discussions.
MW was supported by the Simons Foundation via the Simons Collaboration
on the Many-Electron Problem. EG was supported by DOE grant no. ER46932.
This research used resources of the National Energy Research Scientific
Computing Center, a DOE Office of Science User Facility supported
by the Office of Science of the U.S. Department of Energy under Contract
No. DE-AC02-05CH11231.
\end{acknowledgments}

\bibliographystyle{apsrev4-1}

\end{document}